\documentclass[12pt]{article}
\usepackage{graphics}

\begin{document}
\begin{titlepage}
\begin{flushright}
OUTP-98-P-\\
\end{flushright}

\begin{center}
{\large \bf The D-gauge: a solution to the i-r problem for fermion mass
 generation  in $QED_3$ in the Matsubara formalism}\\
\ \\
by\\
\ \\
D.J. Lee\\
\ \\
Department of Physics\\
Theoretical Physics\\
1 Keble Road\\
Oxford OX1 3NP\\
U.K.\\
\ \\
\ \\
\ \\
\vspace{0.05in}
{\bf Abstract}
\vspace{0.05in}
\end{center}
{\small A serious problem with the Schwinger-Dyson approach to dynamical 
mass generation in $QED_3$ at finite temperature is that the contribution 
from the transverse part of the photon propagator, in the landau gauge, 
leads to infrared divergences in both the mass function $\Sigma$  and the wavefunction renormalization $A$. We show how, by using a simple choice of vertex ansatz and a choice of non-local gauge (the `D-gauge'), both $A$ and $\Sigma$ can be made i-r finite. We formulate an equation for the physical
 mass ${\cal M}=\Sigma/A$, and show that it reduces to the 
corresponding equation obtained in the Landau gauge in the constant 
physical mass approximation ${\cal M}={\cal M}({\bf 0},\pi T)$
(which is finite). Therefore at finite temperature, we are able to justify
a `constant' mass approximation for ${\cal M}$, and show that the value of 
$r={2{\cal M}(T=0) \over k_BT_c}$ remains close to the value obtained in previous calculations which included retardation.}
\vspace{0.05in}
\ \\
\ \\
\begin{flushleft}
May 1998\\
\end{flushleft}
\end{titlepage}

\begin{center}
{\bf 1. Introduction}
\end{center}

A considerable amount of work has been done on investigating fermion mass generation in $QED_3$ both at zero, [1]-[9], and finite temperature [10]-[16].

At finite temperature, one finds a critical temperature, $T_c$, above which the generated fermion mass $\Sigma$ is always zero. An important parameter in the theory is $r$, which is the ratio of $2 \Sigma$ (evaluated at zero 3-momentum and temperature) to $T_c$. This quantity has been calculated in several finite temperature calculations [10]-[16], and turns out to have a value significantly higher than that found in 4-fermion models of B.C.S type, where $r$ is approximately equal to $3.5$. Thus $r$ can serve as an indication of the pairing mechanism (gauge interaction or 4-fermion) in possible applications of $QED_3$ to planar superconductivity.

So far, evidence has been found to support the idea that retardation causes a significant reduction in $r$, unlike other refinements which have been attempted.
In [10], where only the longitudinal part of the  non-retarded propagator was used in the constant mass approximation (i.e assuming the mass is independent of $3$-momentum), $r$ was found to be roughly around $10$. When [10] was refined by the introduction of momentum dependence in the mass [11], there was seen to be little change in $r$. In [12] it was found again that the value of $r$ stayed roughly the same when wave function renormalization was introduced. However, when retardation was introduced, in the calculations of [13] and [14], this value fell to a value of $r \approx 6$. On the introduction of frequency dependence in addition to retardation, [14], it was seen that $r$ only changed by a small amount. Both [13] and 
[14], however, suffered from (different) major technical problems.
 
In [13] (the first calculation to be done taking account of retardation), where  the real time formalism was used, the main difficulty was the non-analyticity of the vacuum polarization at the origin in $k$-space. In 
[13] such behaviour was only able to be treated crudely.

In [14] the Matsubara formalism was used. The major problem for such calculations in the Landau gauge is that there is an i-r divergence in the equation for $\Sigma$ which comes from the transverse part of the photon propagator; so [14] neglected this term. In [16] it was shown that the introduction of a fermion mass in the calculation of $ \Pi_{\nu \mu}$, does not get rid of this divergence.

In the real time formalism a more refined calculation has recently been attempted [15] which tries to improve on [13], with the introduction of a fully $3$-momentum dependent mass. Here, the polarization is treated properly, but this leads to substantially more computational complexity, for no analytic form is found to approximate it in [15]. In these calculations $r$ is found to increase with $N$, the number of flavours of fermion, with the value ranging from $ r \approx 5.4$ (for the $N=1$ case) to $r \approx 14$ (for the $N=3$ case). The value given for $N=1$ is in good agreement with [14], but the results of [14] suggest  a slight reduction in $r$ with increasing $N$, not an increase. Inspection of the data of [15] suggests, however, that some caution may be necessary in accepting the results of the $N=2$ and $3$ cases.

It is our view that the best way to proceed is by using the Matsubara formalism, due to the fact that the non-analyticity problem does not arise. A simple closed form for the photon propagator to leading order in $1/N$ can be found [14]; and this leads to simpler equations than those found in the real time formalism, significantly reducing computational complexity. However, if one is to use the Matsubara formalism one must get round the problem of the i-r divergence. In [16] the divergence was regulated by introducing an arbitrary cutoff to take account of terms beyond leading order in $1/N$ in the vertex, which might regulate the equation for $\Sigma$. In this paper, we indeed see that by going beyond leading order in $1/N$ in the vertex we are able to get a finite equation for${\cal M}=\Sigma/A$ (with the inclusion of the transverse part of the photon propagator), where $A(p_f)$ is the wave function renormalization, without the introduction of an arbitrary cutoff.

We shall divide the work in to four sections and an appendix. In the next section our interest lies in the formulation of equations for both $\Sigma(p_f)$ and $A(p_f)$ at finite temperature for a completely general covariant gauge. If the gauge fixing function is a function of momentum, as it will be in Section.4, then this corresponds to a non-local gauge as discussed in [8] and [9]. The derivation of these equations depends on a simple choice of vertex ansatz for which $\Gamma^{\nu}(p_f,k_f)=A(k_f) \gamma^{\nu} $, which, although it does not satisfy the full Ward-Tahashi identities, is thought to be good starting approximation for $ \Gamma^{\nu}$.

In Section.3, we go on to look at the Landau gauge. Here, we clearly see the problems associated with such a choice of gauge in the equations for $\Sigma(p_f)$ and $A(p_f)$; the poles in the kernals of both equations give rise to logarithmic divergences. Although this is the case, our next step is to see whether we can formulate a finite equation solely in terms of ${ \cal M}$. Indeed, we find such an equation if we make the constant mass approximation 
$ {\cal M}={\cal M}({\bf 0},\pi T)$. However, if we choose our constant mass approximation to have the value of ${\cal M}(p_f)$ at some generally specified point $p'_f=({\bf p'},p'_{0f})$, i.e ${\cal M}={\cal M}(p'_f)$, when ${\bf p'} \neq 0$ we see that the equation for ${\cal M}$ contains divergences.

In Section.4 we show that for a choice of non-local gauge, which we shall call the $D$-gauge, the equations for $A(p_f)$ and $ \Sigma(p_f) $ are both finite. Again we shall be able to derive an equation solely in terms of ${ \cal M}(p_f)$. By making the constant mass approximation $ {\cal M}={\cal M}({\bf 0}, \pi T)$ we are able to show that this equation reduces to the same equation as in the Landau Gauge (making the same constant mass approximation), if we use the form
of the photon propagator developed in [14]. However, there are two advantages to using the D-gauge. The first is that, unlike the Landau gauge, if we choose our constant mass to be $ {\cal M}={\cal M}(p'_f)$, where $p'_f$ is our generally specified point, we do not encounter any divergences. The upshot of this is that our equation for  $ {\cal M}={\cal M}({\bf 0}, \pi T)$ can be considered a far more sensible approximation in this gauge. Furthermore, if one introduces full $3$-momentum dependence into ${\cal M}(p_f)$, the equation for ${\cal M}$ remains finite. The second advantage is that at $T=0$ this gauge is identical to the Landau gauge.

In the last section, Section.5, we solve the equation for the constant mass approximation $ {\cal M}={\cal M}({\bf 0}, \pi T)$. From our numerical solution we are able to calculate a value of $r$, which we find roughly to be $r \approx 5.5$. When we compare these results with those of [14], we see that these results compare favourably to those of [14] (using the constant mass approximation) with the number of flavours halved, the values of $\Sigma(T=0)$
being in good agreement.

Finally in the appendix we prove an important result in our analysis.

\begin{center}
{\bf 2. The Schwinger-Dyson equation}
\end{center}

For massless $QED_3$ with $N$-flavours we have the usual Lagrangian density:
 \begin{equation}
{\cal L} = -1/4 f_{\mu \nu}f^{\mu \nu}+ \sum_{i} \bar{\psi}_i (i \partial \!\!\! / -e a \!\!\! /) \psi_i 
\end{equation}
where $a_\mu$ is the vector potential and $i=1,2...N$. As in previous work ([14] and [16] ) we have chosen a reducible representation for the Dirac algebra, so that (1) has continuous chiral symmetry as discussed in [2]. The full Schwinger-Dyson (S-D) equation for the fermion propagator takes the form (in the Matsubara formalism)
\begin{equation}
S^{-1}_{F}(p_f)= S^{(0)-1}_F(p_f)-{e \over \beta} \sum^{\infty}_{n=-\infty} \int {d^2k \over (2 \pi)^2} S_F(k_f) \gamma^{\mu} \Delta_{\mu \nu} (k_f-p_f) \Gamma^{\mu}(k_f-p_f,k_f).
\end{equation}
In equation (2) we use (as in [14] and [16]) the subscript $f$ to denote fermionic 3-momenta, with $k_f=(k_{0f},{\bf k})$; the zeroth component has the form $k_{0f}={2 \pi (m+1/2) \over \beta}$ in the Matsubara formalism, where $m$ is an integer. To denote bosonic momenta we shall use the subscript $b$, with $p_b=(p_{0b},{\bf p})$; the zeroth component is of the form $p_{0b}= {2 \pi m \over \beta}$, where $m$ is again integer.

In this paper our interest lies in going beyond leading order in $1/N$ in the vertex part. Our main motivation for this stems from the work of [14] and [16], in which it was shown that if the bare vertex is used, the transverse part of the photon propagator leads to an i-r divergence in the S-D equation at finite temperature. Furthermore, it was shown in [16], that the introduction of a dynamically generated fermion mass, $\Sigma$, into the fermion propagators in the calculation of the vacuum polarization, $\Pi_{\mu \nu}$ did not alter this conclusion. The main hope of [16] was that if one went beyond leading order in $1/N$, such i-r divergences would be regulated, without the inclusion of an arbitrary cutoff (c.f [16]). 

Let us first write down a form for the fermion propagator, which takes account of wavefunction renormalization
\begin{equation}
S^{-1}(k_f)=A(k_f)p \! \! \! / + \Sigma(k_f)
\end{equation}
where $A(k_f)$ is the wavefunction renormalization and $\Sigma$ is the fermion mass function. Now since we interested in going beyond leading order in $1/N$ in the vertex, we no longer take $\Gamma^{\nu}$ to be its bare value of $e \gamma^{\nu}$; instead we choose $\Gamma^{\nu}$ to be an ansatz of the form:
\begin{equation}
\Gamma^{\nu}(k_f-p_f,k_f)=A(k_f)e \gamma^{\nu}.
\end{equation}
Here one should note that although (3) may be considered as an exact expression for $S^{-1}(k_f)$, the expression for $\Gamma^{\nu}$ does not obey the Ward-Tahashi identities relating $S^{-1}(k_f)$ to $\Gamma^{\nu}(k_f-p_f,k_f)$.
Therefore (4) can only be considered at best an approximation of the true vertex. However, at zero temperature the vertex function is thought to contain only terms of either zeroth or first order in $A(p_f)$ (c.f [3] and [7]); our ansatz is thought to be a reasonable starting approximation to make for the vertex function. 

In our study we shall neglect terms beyond leading order in $1/N$ that contribute to the full photon propagator (these terms are thought to contribute little to the result (c.f [3]) at zero temperature); therefore we shall be using the approximate form for the full photon propagator (to leading order in $1/N$) in the numerical calculations of the fermion mass, first developed in [14]. For the moment, however, we shall write down a more general form for the full photon propagator:
\begin{equation}
\Delta_{\mu \nu} (q_b)= D_L(q_b) A_{\mu \nu} + D_T(q_b) B_{\mu \nu}+ {\epsilon(q_b) \over q^2_b } {q_{\mu b} q_{\nu b} \over q^2_b}
\end{equation}
where $A_{\mu \nu}$ and $B_{\mu \nu}$ are the longitudinal and transverse projection operators, respectively. $\epsilon(q_b)$ is a general covariant gauge fixing function, which in general, we make dependent on 3-momentum. If $\epsilon (q_b)$ is not a constant w.r.t $q_b$, then one has a non-local gauge, as discussed in [8] at zero temperature.

 From (3),(4) and (5) we are able to write the S-D equation (fig.1) in the form
\begin{equation}
S_F(P_f) \equiv A(p_f) p \!\!\! / + \Sigma(p_f) I = p \!\!\! / - {\bf \Sigma}(p_f)
\end{equation}  
where 
\begin{equation}
{\bf \Sigma} (p_f)= {e^2 \over \beta} \sum_n \int {d^2k \over (2 \pi)^2} {\gamma_{\mu} \gamma_{\nu} A(k_f) \over k \!\!\! /_f A(k_f) +B(k_f)} \left[ D_L(q_b) A_{\mu \nu}+ D_T(q_b) B_{\mu \nu} + {\epsilon(q_b) \over q^2_b }{q_{\mu b} q_{\nu b} \over q^2_b} \right]. 
\end{equation}
If we take the trace of (7), we get the following equation for $\Sigma(p_f)$:
\begin{equation}
\Sigma(p_f)= {e^2 \over \beta} \sum_n \int {d^2k \over (2 \pi)^2} { {\cal M} (k_f) \over k_f^2 + {\cal M}^2(k_f)} \left[ D_L+D_T+{\epsilon(q_b) \over q^2_b} \right]
\end{equation}
where $ {\cal M}(p_f)= {\Sigma(p_f) \over A(p_f)}$.

By multiplying by $p \!\!\! /$ and taking the trace we are able to derive the following expression for $A(p_f)$:
\begin{eqnarray}
A(p_f)=1+{1 \over p^2_f}{e^2 \over \beta} \sum_n \int {d^2k \over (2 \pi)^2} {1 \over k^2_f+ {\cal M}^2(k_f)} \left[ (p_f.k_f) \left( D_L(q_b)+D_T(q_b)+ {\epsilon(q_b) \over q^2_b} \right) \right. \nonumber \\
\left. -2T_1(p_{0f},k_{0f},{\bf p},{\bf k})D_L(q_b)
-2T_2({\bf p},{\bf k})D_T(q_b)-2 {\epsilon(q_b) \over q^4_b}(p_f.q_b)(k_f.q_b) \right]
\end{eqnarray}
\begin{eqnarray}
\mbox{ where } T_1 &= &p^{\mu}_f k^{\nu}_f A_{\mu \nu} \nonumber \\
                   &= & \left( p_{0f}-{(q_b.p_f)q_{0b} \over q^2_b}
\right){q^2_b \over {\bf q}^2} \left( k_{0f}- {(q_b.k_f)q_{0b} \over {\bf q}^2}
\right) \\
\mbox{ and } T_2 &= &p^{\mu}_f k^{\nu}_f B_{\mu \nu}= {\bf k.p}-{ {\bf k.q}{\bf k.q}
\over {\bf q}^2}. 
\end{eqnarray}
Now that we have two equations (`` gap equations '') for both $\Sigma$ and $A$, the  step is to investigate whether or not these equations are i-r finite, and to obtain an equation solely in terms of the physical mass, ${\cal M}(p_f)$, in the Landau gauge. This we investigate in the next section.

\begin{center}
{\bf 3. The gap equations in the Landau gauge}
\end{center}

We shall first consider the Landau gauge, setting $\epsilon(q_b)=0$. The contribution to the R.H.S of (8) most likely to contain i-r divergences is the $p_{0f}=k_{0f}$ mode:
\begin{equation}
\Sigma^0(p_f) ={e^2 \over \beta} \int {d^2k \over (2 \pi)^2} {{\cal M}({\bf k},p_{0f}) \over {\bf k}^2+p^2_{0f}+{\cal M}^2({\bf k},p_{0f})}[D^0_L({\bf q})+D^0_T({\bf q})]
\end{equation}
where $D^0_L( {\bf q})$ and $D^0_T( {\bf q})$ are the $q_{0b}=0$ contributions to $D_L$ and $D_T$, respectively. Already, we know from the work of [14] and
 [16], that in the approximation $A(p_f)=1$, $\Sigma=\Sigma( {\bf 0}, \pi T )$ (which we call the constant mass approximation ), $\Sigma^0$ contains an i-r divergence at $k_{0f}= \pi T$. We now see whether a divergence exists for all $p_f$ at $p_f=k_f$.
   From the results of [10] and [14], we see that for small ${\bf q}$
\begin{eqnarray}
D^0_T({\bf q}) & \propto & {1 \over ({\bf k}-{\bf p})^2} \\
\mbox{ and } D^0_L({\bf q}) & \propto & {1 \over ({\bf k}-{\bf p})^2+M^2_p}
\end{eqnarray}
where $M_p= \sqrt{{ 2 \alpha \ln 2 \over \pi \beta}}$. Already we see that any potential divergence must come from the $D^{0}_T$ contribution. By replacing $D^0_T$ by its small ${\bf q}$ behaviour and neglecting $D^0_L$ in (12), we have on performing the angular integration:
\begin{equation}
\Sigma^0_{div}= {Ce^2 \over \beta} \int { |{\bf k}| d|{\bf k}| \over (2 \pi)} 
{ { \cal M}({\bf k},p_{0f}) \over {\bf k}^2+p_{0f}^2+{\cal M}^2({\bf k},p_{0f})}
{1 \over |{\bf k}^2-{\bf p}^2|}
\end{equation}
where $C$ is a constant of proportionality. One can see that (15) is manifestly divergent at ${\bf k}={\bf p}$. An important point to note, is the appearance of a factor $1/\beta$ in front of the integral. This factor ensures that at $T=0$ there are no divergences in the equation for $\Sigma$, which is known to be finite at $T=0$.

From (15) and by comparison of (8) and (9) we see that $A(p_f)$ contains the same type of divergences. Although $A(p_f)$ and $\Sigma(p_f)$ both contain divergences, the question to ask is: can we find a finite equation for the physical mass ${\cal M}(p_f)$, in which the divergences cancel? The answer is, indeed, that we can find an equation solely in terms of ${\cal M}(p_f)$; however this equation generally contains divergences.

To find an equation for ${\cal M}(p_f)$, let us rewrite the L.H.S of (8) as $A(p_f) {\cal M}(p_f)$ and then use equation (10) to eliminate $A(p_f)$, leaving us with
\begin{eqnarray}
{\cal M}(p_f)={e^2 \over \beta} \sum_n \int {d^2k \over (2 \pi)^2} \left[ { {\cal M}(k_f) (D_L(q_b)+D_T(q_b)) \over k_f^2+{\cal M}^2(k_f)+k_f^2 } \right.
\nonumber \\
  \left. - {{\cal M}(p_f) p_f^{-2} ((p_f.k_f)(D_L(q_b)+D_T(q_b))-2T_1(p_{0f},k_{0f},{\bf p},{\bf k})D_L(q_b)-
2T_2({\bf p},{\bf k})D_T(q_b))  \over k_f^2+{\cal M}^2(k_f)+k_f^2 } \right].
\end{eqnarray}
Now, our first goal would be to solve this equation for a constant mass 
${\cal M}={\cal M}(p_f)$. Instead of specifying the physical mass at the point $({\bf 0},\pi T)$ for all $p_f$ (as is usually done in similar calculations), let us choose ${\cal M}$ to be the same as the value of ${\cal M}(p_f)$, specified
at some general point $p'_f=(p'_{0f},{\bf p'})$ for all $p_f$. In this approximation, ${\cal M}={ \cal M}(p'_f)$, equation (16) becomes
\begin{eqnarray}
1={e^2 \over \beta} \sum_n \int {d^2k \over (2 \pi)^2} {1 \over {\cal M}^2+k_f^2} \left[ D_L(k_f-p'_f)+D_T(k_f-p'_f) -{1 \over p'^2_f } [(p_f'.k_f) (D_L(k_f-p'_f) \right. \nonumber \\
 \left.  
+D_T(k_f-p'_f))-2T_1(p'_{0f},k_{0f},{\bf p}',{\bf k})D_L(k_f-p_f')-2T_2({\bf p}',{\bf k})
D_T(k_f-p'_f)] \right].
\end{eqnarray}
Possible danger comes from the $k_f=p'_f$ pole in (17), so we look at the contribution to the R.H.S of (17) where $k_{0f}=p'_{0f}$, which we call 
${\cal M}^0$.
\begin{eqnarray}
{\cal M}^0={e^2 \over \beta p^2_f} \int {d^2 k \over ( 2 \pi)^2 }
{1 \over {\cal M}^2+p'^2_{0f}+{\bf k}^2} \left(
{\bf p}'^2-{\bf p}'.{\bf k}+2p^2_{0f})D^0_L({\bf k}-{\bf p}') \right.
\nonumber \\
\left. \left( {\bf p'.k}-{2 {\bf k.(k-p')}{\bf p.(k-p')} \over ({\bf k}-{\bf p})^2 }+{\bf p}'^2 \right)D^0_T({\bf k}-{\bf p}') \right).
\end{eqnarray}
Terms in ${\cal M}^0$ proportional to $D^0_L({\bf k}-{\bf p'})$ are finite, for they are regulated by $M^2_p$ in the denominator (c.f. (14)). We are also able to show that the contribution from
$$ \left({ \bf p.k } -{2 {\bf k.(k-p')}{\bf p.(k-p')} \over ({\bf k}-{\bf p})^2 } \right) D^0_T({\bf k}-{\bf p}')  $$
is finite (see Appendix), although at ${\bf k}={\bf p}'$ there exists a term proportional to a delta function which vanishes as ${\bf p}' \rightarrow 0$. This leaves us with a term proportional to ${\bf p}'^2 D^0_T({\bf k}-{\bf p})$,
which is divergent when we choose ${\bf p}' \neq {\bf 0}$.
   
The upshot of the above is that although we have been able to find a constant mass equation which is finite at ${\bf p}'=0$, this approximation is not a sensible one to make, due to the wild discrepancies between points where 
${\bf p}'={\bf 0}$ and those where ${\bf p}' \neq {\bf 0}$ (the equation for
${\bf p}' \neq {\bf 0}$ leads to a singular mass). These singularities are also manifest if one decides to relax the constant mass approximation. In the next section we shall work in a gauge which, as we shall see, removes these singularities.  
 
\begin{center}
{\bf 4. The $D$-gauge}
\end{center}

If we choose a non-local gauge $\epsilon(q_b)=(D_L-D_T)q^2_b$ (which we call the D-gauge) then, already, we 
see from (8) that
\begin{equation}
\Sigma(p_f)={2e^2 \over \beta} \sum_n \int {d^2 k \over (2 \pi)^2} {{\cal M}(k_f) \over k^2_f+ {\cal M}(k_f)} D_L(q_b),
\end{equation}
which is finite for all $p_f$. In this gauge we are also able to rewrite
(9) as
\begin{eqnarray}
A(p_f)=1+{1 \over p^2_f} {e^2 \over \beta} \sum_n \int {d^2 k \over (2 \pi)^2}
{1 \over k^2_f + {\cal M}^2(k_f)} \left[ 2(p_f.k_f)D_L(q_b) \right.\nonumber \\
\left. -2T_1(p_{0f},k_{0f},{\bf p},{\bf k})D_L(q_b)
-2T_2({\bf p},{\bf k})D_T(q_b)+{2(p_f.q_b)(k_f.q_b) \over q_b^2}(D_T(q_b)-D_L(q_b))
\right]. 
\end{eqnarray}
Again the contribution most likely to contain divergences in the R.H.S of (20) is the $k_{0f}=p_{0f}$ contribution.
\begin{eqnarray}
A^0(p_f)={1 \over p^2_f}{e^2 \over \beta} \int {d^2 k \over (2 \pi)^2} 
{1 \over {\bf k}^2+p^2_{0f}+ {\cal M}({bf k},p_{0f}) } \nonumber \\
\times 2  \left( \left[ ({\bf p}.{\bf k})-{({\bf p.q})({\bf k.q}) \over {\bf q}^2}
\right] D^0_L({\bf q})- 2 \left[ ( {\bf p}.{\bf k})- {2({\bf p.q})({\bf k.q}) \over {\bf q}^2} \right] D^0_T({\bf q}) \right).
\end{eqnarray}
Since we already know that the term
$$ \left({ \bf p.k } -{2 {\bf k.(k-p')}{\bf p.(k-p')} \over ({\bf k}-{\bf p})^2 } \right) D^0_T({\bf k}-{\bf p}')  $$
gives a finite contribution (see appendix), it is easy to see that (21) is i-r finite. The consequence of this is that we have a gauge in which our equations for $\Sigma$ and $A$ are finite. An advantage of this gauge (as opposed to other gauges we might choose to cancel this divergence) is that in the limit $T \rightarrow 0$, this gauge joins smoothly on to the Landau
gauge, for at $T=0$ $D_L=D_T$. So at $T=0$, this gauge has exactly the same gap equations as those of the Landau gauge, which has been extensively studied.

From (19) and (20) we are able to derive an  equation solely in terms of the physical mass, ${\cal M}$:
\begin{eqnarray}
{\cal M}(p_f)= {e^2 \over \beta} \sum_n \int {d^2k \over (2 \pi)^2} \left[ {2 {\cal M}(k_f) D_L(q_b) \over {\cal M}^2(k_f)+k^2_f} \right. \nonumber \\
 \left. -{1 \over p^2_f}{ {\cal M}(p_f) \over {\cal M}^2(p_f)+k^2_f} \left[ 2(p_f.k_f)
D_L(q_b)-2T_1(p_{0f},k_{0f},{\bf p},{\bf k})D_L(q_b) \right. \right. \nonumber \\
\left. \left. -2T_2({\bf p},{\bf k})D_T(q_b)+{2(p_f.q_b)(k_f.q_b) \over q^2_b}
(D_T(q_b)-D_L(q_b)) \right] \right].
\end{eqnarray}
This equation, unlike (16), is finite for all $p_f$, due to the fact that both (19) and (20) are finite. Again, as in the previous section, we make the constant mass approximation, now choosing $p_f'=({\bf 0}, \pi T)$. In this approximation (22) reduces to
\begin{eqnarray}
1={e^2 \over \beta} \sum_n \int {d^2k \over (2 \pi)^2} \left[ {2 D_L(k_b) \over {\cal M}^2+k_f^2}- \left[ \left( {1 \over (T \pi)^2} (2 \pi T k_{0f}) D_L(k_b)
 \right. \right. \right. \nonumber \\
\left. \left. \left. -2 (\pi T)^2  \left( 1- {k^2_{0b} \over k^2_b}\right) 
D_L(k_b) \right) + 
{ 2 \pi T k_{0b} k_f.k_b \over k_b^2} (D_T(k_b)-D_L(k_b)) \right] \left( {1 \over {\cal M}^2+k_f^2} \right)  \right]. 
\end{eqnarray}
Now using the approximation for the photon propagator given in [14], (23) becomes:
\begin{eqnarray}
1={ \alpha \over N \beta } \sum_{n \neq 0} \int {d^2 k \over (2 \pi)^2 } \left(
{4 D^n(k_b) \over {\cal M}^2+k^2_f } - {2 k_{0f} \over \pi T} {D^n(k_b) \over {\cal M}^2+k_f^2} \right. \nonumber \\
\left. -{2 k_{0b}^2 \over k^2_b } {D^n(k_b) \over {\cal M}^2+k_f^2} \right)
+{\alpha \over N \beta } \int {d^2 k \over (2 \pi)^2} {2 D^0_L(k_b) \over {\cal M}^2+k^2_f}
\end{eqnarray}
where $\alpha= Ne^2$
\begin{eqnarray}
D^n(k_b)={1 \over k_b^2+ \alpha k_b/8} \; \;  k_{0b} \neq 0, \\
D^0_L({\bf k})= { 1 \over {\bf k}^2+ \Pi^3_0({\bf k}) } \; \; k_{0b}=0, \\
\mbox{ and } \Pi^0_3( {\bf k})= {1 \over 8} \left( \alpha \over \beta \right)
\left[ |{\bf k}| \beta + {16 \ln 2 \over \beta} exp \left( - \pi \beta |{\bf k}|
\over 16 \ln 2 \right) \right].
\end{eqnarray}
One can easily check that, in the Landau gauge, using the same constant mass approximation ( ${\cal M}={\cal M}(p'_{0f},{\bf p}')={\cal M}(\pi T, {\bf 0})$)
and the same form for the full photon propagator, (17) leads to the same equation as (24). However, in this non-local gauge (24) seems a more sensible 
equation to solve, due to (22) having no divergences for any $p_f$. So if we wanted to solve (22) with a constant physical mass approximation, specified at some other point, say ${\cal M}={\cal M}(p'_f)$, we would still get a finite result for ${\cal M}$. This gauge has an added advantage; if we wanted to solve (22) with full $3$-momentum dependence, we would still get sensible results.

\begin{center}
{\bf 5. The numerical solution to the constant physical mass equation (24)}
\end{center}

By following similar steps to those of [14], we are able to recast (24) in a form convenient for numerical analysis
\begin{eqnarray}
1  = {a \over \pi N} \sum^{\infty}_{n=1} \left[
(1-2n)I_1(2 \pi n,0.125a,(a^2s^2+(2 \pi (n+1/2))^2)^{1/2}) \right. \nonumber \\
 \left.  +  (1+2n)I_1(2 \pi n, 0.125a, (a^2s^2+(2 \pi (n-1/2))^2)^{1/2}) \right. \nonumber \\
\left.   -  (2 \pi n)^2(I_2(2 \pi n,0.125a,(a^2s^2+(2 \pi (n+1/2))^2)^{1/2}) \right. \nonumber \\
 \left.  +  I_2( 2 \pi n,0.125a,(a^2s^2+(2 \pi (n+1/2))^2)^{1/2}) \right]
\nonumber \\
 +  {a \over \pi N} \int^{\infty}_{0} dx {x \over x^2+ \beta^2 \Pi^3_0(x)}{1 \over x^2+a^2s^2+\pi^2}
\end{eqnarray}
where $s={\cal M}/\alpha$, $a=\alpha \beta$ and $x= \beta |{\bf k}|$; also
\begin{equation}
I_1(d,a,c)={1 \over 2(a^2+c^2-d^2)} \ln \left( {c^2 \over (d+a)^2} \right)
+{a \over (c^2-d^2)^{1/2}(a^2+c^2-d^2)} \arctan \left( {(c^2-d^2)^{1/2} \over d}
\right) 
\end{equation}
and
\begin{eqnarray}
I_2(d,a,c)={1 \over a(c^2-d^2)d}+\left( { (a^2+c^2-d^2) \ln d -(c^2-d^2) \ln (d+a)-a^2 \ln c \over a^2(c^2-d^2)(a^2+c^2-d^2) } \right) \nonumber \\
-{a \over (a^2+c^2-d^2)(c^2-d^2)^{3/2}} \arctan \left( (c^2-d^2)^{1/2} \over a \right).
\end{eqnarray}
Before proceeding to solve (28) numerically we require the zero temperature limit of (24). This is most easily found by going back to (22), due to the fact that the 
sums in (28) and (24) cannot be performed analytically. In the $T \rightarrow 0$ limit (22) becomes:
\begin{equation}
p^2{\cal M}(p)=\int {d^3k \over (2 \pi)^3} \left[ {2p^2 {\cal M}(k) \over 
{\cal M}^2(k)+k^2}-{2 (p.q)(q.k) \over q^2}{ {\cal M}(p) \over {\cal M}^2(k)+k^2} \right] \left( {1 \over q^2+\alpha q/8} \right).
\end{equation}
By Taylor expanding (31) in powers of 3-momentum, $p$, keeping only terms up to order $p^2$ and performing the angular integration, we are able to get the following expression for $s=\cal{M}/\alpha$ in the $p \rightarrow 0$ limit:
\begin{equation}
s(0)={2 \over (2 \pi)^2 N} \left[ \int^{\infty}_0 dy {2s(y) \over (y+0.125)
(y^2+s^2(y))}+{1 \over 12} \int^{\infty}_0 dy {s(0) \over (y+0.125)^2 (y^2+s^2(y))} \right]
\end{equation}
where $y=k/\alpha$. Since $S(y)$ is monotonically decreasing, so as to satisfy (22), it is true in general that
\begin{eqnarray}
s(0) \ge  {4 \over (2 \pi)^2 N} \int^{\infty}_0 dy {s(y) \over (y+0.125)(y^2+s^2(y))} \nonumber \\
 \left( 1-{1 \over 6(2 \pi)^2N} \int^{\infty}_0 dy {1 \over (y+0.125)^2+(y^2+s^(0))} \right)^{-1}.
\end{eqnarray}
If we look at (33) we see that the effect of wavefunction renormalization is to increase $s(0)$, for if we set $A=1$ we neglect the second term in the R.H.S of (32). In the constant mass approximation $s(y)=s(0)$ this bound becomes saturated: (33) becomes an equation for $s(0)$. On performing the integrals in the constant mass approximation (33) reduces to:
\begin{eqnarray}
1={4 \over (2 \pi)^2 N} [F_1(s)+F_2(s)/3] \\
\mbox{ where } F_1= \left( 1-{(0.125)^2 \over (0.125)^2+s^2} \right) {\pi \over 2s}-{0.125 \over (0.125)^2+s^2} \ln \left( {s \over 0.125} \right) \\
\mbox{ and } F_2= \left( {(0.125)0.25 \over (0.125)^2+s^2}-{0.25(0.125)^2 \over ((0.125)^2+s^2)^2} \right) \left( \pi \over 2s \right) \nonumber \\
+ \left( {0.125 \over (0.125)^2+s^2} -{(0.125)^2 0.25 \over ((0.125)^2+s^2)^2} 
\right) \ln \left( {s \over 0.125} \right) \nonumber \\
-{0.125 \over (0.125)^2+s^2}.  
\end{eqnarray}
Now we are able to plot graphs of the solutions as functions of $(k_B T/\alpha)$ for various $N$ (see fig.2), by solving (28) and (34) numerically using Mathematica. If one halves the number of flavours given for each solution in fig.1, these results differ little from those of [14], the values of $\Sigma$ being in good agreement. In fig.3 we show a graph of $T_ck_B/ \alpha$, the critical temperature plotted against N. As expected, we see that as $N$ increases $T_ck_B$ falls. However, there is no finite $N_c$, a critical number of flavours. This is due to the singular behaviour of the R.H.S of (34) at $s=0$, from which we infer that $s=0$ is not a solution of (34) at a finite value of $N$; therefore $N_c \rightarrow \infty$, as was shown in [14] and [16] for the $A=1$, $\Sigma=\Sigma(\pi T,0)$ approximation. There is strong evidence to support the existence of a finite $N_c$ in the full, untruncated mass gap equations in the absence of any approximations. The reason for this 
discrepancy is that our approximations are cruder than those used at $T=0$. Even though this is the case, we expect the exact solution (at the point $(\pi T, {\bf 0})$) to have, at least, the same qualitative features as those shown in fig.2.

Finally we present a table of values (Table.1) for 
$r={2 {\cal M} \over k_bT_c}$,
which changes little with $N$ and takes the average value $r=5.56$. When compared to the results of [13] and [14], it is evident that wavefunction renormalization does little to change the value of $r$. The same conclusions were found when [12] was compared with [11] in the unretarded case.

An obvious extension to this work would be to relax the constant mass approximation by introducing frequency or momentum dependence (or both) into our equation for the physical mass. Another possible extension would be to use a more sophisticated vertex ansatz. A good one to choose would be the Ball-Chu ansatz which obeys the Ward identities at zero temperature. The main difficulty is that the form of the equations for $A$ and $\Sigma$ would be far more complicated than those encountered in this paper. As this may well be, one long term goal should be to use an ansatz which respects the Ward identities at finite temperature, which would reduce to the Ball-Chu ansatz at zero temperature (the Ball-Chu ansatz does not respect the Ward identities at finite temperature, due to loss of Lorenz invariance). Such an ansatz would get rid of the divergences seen in the Landau gauge and would produce a gauge dependent physical mass. Until then the $(D_L-D_T)q^2_b$ gauge is a sensible choice of gauge to work in for two reasons already stated: firstly this equation leads to finite equations for $A(p_f)$ and $\Sigma(p_f)$, secondly at $T=0$ this gauge is identical to the Landau gauge.

From this paper, one can plainly see that there is still a considerable amount of work to be done at finite temperature in $QED_3$, especially if we want our results to be consistent with new results at zero temperature ([7],[8] and [9]).

\setcounter{equation}{0}
\renewcommand{\theequation}{\mbox{A.\arabic{equation}}}

\begin{center}
{\bf Appendix}
\end{center}

In the appendix we shall show that the term
\begin{equation}
\left( {\bf p.k}-{2{\bf k.q}{\bf q.p} \over {\bf q}^2} \right)D_T^0({\bf q}={\bf k}-{\bf p})
\end{equation}
in (21) leads to a finite integral when ${\bf k-p} \rightarrow {\bf 0}$, although we shall see that a delta function contribution exists at $ {\bf k-p}={\bf 0}$. To show this we consider the integral
\begin{equation}
I= \int^{2 \pi}_0 {d \theta \over 2 \pi} \left(
{\bf p.k}-{2{\bf k.q}{\bf q.p} \over {\bf q}^2} \right) {1 \over {\bf q}^2}.
\end{equation}
In (A.2) we have replaced $D^0_T({\bf k-p})$ by its leading order behaviour in ${\bf q}$. This is sufficient, for we are interested whether or not there is any divergent behaviour at ${\bf p}={\bf k}$; higher order corrections do not matter, for they are unable to cause divergent behaviour when integrated over ${\bf k}$ (the next highest term is proportional to $1/|{\bf q}|$). We are able to write (A.2) in terms of a complex integral (where we have introduced a regulator $\epsilon$, so that we are able to probe the ${\bf k}={\bf p}$ region.)
\begin{equation}
I={1 \over 2 \pi i} \oint_{|z|=1} \left[ {2 {\bf p}^2 {\bf k}^2 z-({\bf k}^2+
{\bf p}^2)|{\bf k}||{\bf p}|(z^2+1)/2 \over {\bf k}^2 {\bf p}^2 (z-|{\bf p}|/|{\bf k}|+i \epsilon/(2 |{\bf k}||{\bf p}|))^2(z-|{\bf k}|/|{\bf p}|-i \epsilon / (2 |{\bf k}||{\bf p}|))^2} \right].
\end{equation}
We are able to evaluate this integral, giving us the following expressions.
\begin{eqnarray}
I= \left[ {2 {\bf p}^2{\bf k}^2-({\bf k}^2+{\bf p}^2)({\bf p}^2-i \epsilon/2)
\over ({\bf p}^2-{\bf k}^2-i \epsilon)^2 }
-{(4 |{\bf p}|^3|{\bf k}|-2i |{\bf p}||{\bf k}| \epsilon)|{\bf p}||{\bf k}| \over  ({\bf p}^2-{\bf k}^2-i \epsilon)^3} \right. \nonumber \\
 \left. +{ ({\bf k}^2+{\bf p}^2) {\bf k}^2{\bf p}^2 ( {\bf p}^2/{\bf k}^2 - i \epsilon/ {\bf k^2}-{\epsilon}^2/( 4 {\bf k}^2 {\bf p}^2) +1) \over  ({\bf p}^2-{\bf k}^2-i \epsilon)^3} \right] \mbox{ when } |{\bf p}|<|{\bf k}|(1-\delta^2)^{1/2} \\
I=0 \mbox{ when } |{\bf k}|(1+\delta^2)^{1/2}> |{\bf p}|>  |{\bf k}|(1+\delta^2)^{1/2} \\
\nonumber \\
I= \left[ {2 {\bf p}^2{\bf k}^2-({\bf k}^2+{\bf p}^2)({\bf k}^2+i \epsilon/2)
\over ({\bf k}^2-{\bf p}^2+i \epsilon)^2 }
-{(4 |{\bf k}|^3|{\bf p}|+2i |{\bf p}||{\bf k}| \epsilon)|{\bf p}||{\bf k}| \over  ({\bf k}^2-{\bf p}^2+i \epsilon)^3} \right. \nonumber \\
\left. +{ ({\bf k}^2+{\bf p}^2) {\bf k}^2{\bf p}^2 ( {\bf p}^2/{\bf k}^2 - i \epsilon/ {\bf k^2}-{\epsilon}^2/( 4 {\bf k}^2 {\bf p}^2) +1) \over  ({\bf p}^2-{\bf k}^2-i \epsilon)^3} \right] \mbox{ when } |{\bf p}|>|{\bf k}|(1+\delta^2)^{1/2} 
\end{eqnarray}
where $\delta= {\epsilon \over 2 |{\bf k}||{\bf p}|}$. By checking (A.4) and (A.6) it is easy to show that the terms of zero order in $\epsilon$, in the numerator, vanish in both expressions when $\epsilon \rightarrow 0$. For terms of first order in $\epsilon$ (in the numerator) we have the following expressions
\begin{eqnarray}
I_{\epsilon}= \left[  {-i \epsilon/2 \over ( {\bf p}^2-{\bf k}^2-i \epsilon)}+
{ \epsilon^2/2 \over ( {\bf p}^2-{\bf k}^2- i \epsilon)^2}+
{ {\bf p}^2 \epsilon^2 \over  ( {\bf p}^2-{\bf k}^2- i\epsilon)^3 } \right]
\nonumber \\
\mbox{ when } |{\bf p}|<|{\bf k}|(1- \delta^2)^{1/2} \\
I_{\epsilon}= \left[  {+i \epsilon/2 \over ( {\bf k}^2-{\bf p}^2+i \epsilon)}+
{ \epsilon^2/2 \over ( {\bf k}^2-{\bf p}^2+ i\epsilon)^2}+{ {\bf k}^2 \epsilon^2
\over  ( {\bf p}^2-{\bf k}^2+ i \epsilon)^3 } \right] \nonumber \\
\mbox{ when } |{\bf p}|>|{\bf k}|(1+\delta^2)^{1/2}. 
\end{eqnarray}
In both (A.6) and (A.7) the first two terms cancel when $\epsilon \rightarrow 0$, but the last terms in both expressions is singular at ${ \bf k}^2-{\bf p}^2=0$ (zero everywhere else). When we consider terms that are second order in $\epsilon$ (in the numerator) we see, that also, these terms are singular at ${\bf k}^2-{\bf p}^2=0$.
\begin{eqnarray}
I_{\epsilon^2}=-{({\bf k}^2+{\bf p}^2) \epsilon^2/4 \over ( {\bf p}^2 -{\bf k}^2
-i \epsilon)^3 } \mbox{ when } |{\bf p}|<|{\bf k}|(1- \delta^2)^{1/2} \\
 I_{\epsilon^2}=-{({\bf k}^2+{\bf p}^2) \epsilon^2/4 \over ( {\bf k}^2 -{\bf p}^2+i \epsilon)^3 } \mbox{ when } |{\bf p}|<|{\bf k}|(1- \delta^2)^{1/2}.
 \end{eqnarray}
Combining (A.7) with (A.9) and (A.8) with (A.10) gives us for the singular part of $I$:
\begin{eqnarray}
I_{sing}=[ {\bf p}^2-({\bf k}^2+{\bf p}^2)/4] \left( \epsilon^2 \over (
{\bf p}^2-{\bf k}^2- i \epsilon)^3 \right) \mbox{ when } |{\bf p}|<|{\bf k}|(1- \delta^2)^{1/2} \nonumber \\
I_{sing}=0 \mbox{ when } |{\bf k}|(1+ \delta^2)^{1/2}>|{\bf p}|>|{\bf k}|(1- \delta^2)^{1/2} \nonumber \\
I_{sing}=[ {\bf k}^2-({\bf k}^2+{\bf p}^2)/4] \left( \epsilon^2 \over (
{\bf k}^2-{\bf p}^2+ i \epsilon)^3 \right) \mbox{ when } |{\bf p}|>|{\bf k}|(1+ \delta^2)^{1/2}.
\end{eqnarray}
In the limit $\epsilon \rightarrow 0$, (A.10) suggests that $I_{sing}$ will be proportional to a delta function
\begin{equation}
I_{sing}=({ \bf p}^2/2) {\cal N} \delta({\bf k}^2- {\bf p}^2)
\end{equation}
where we are able to evaluate ${\cal N}$
\begin{eqnarray}
{\cal N} &=& \lim_{\epsilon \rightarrow 0} \left\{ \int^{\infty}_{ \delta^2 {\bf k}^2}
{ \epsilon^2 du \over (u+i \epsilon )^3} -  \int^{- \delta^2 {\bf k}^2}_{- \infty} {\epsilon^2 du \over (u+i \epsilon )^3} \right\} \nonumber \\
&=& \lim_{\epsilon \rightarrow 0} \left\{ {\epsilon^2 \over 2 (\delta^2 {\bf k^2} +i \epsilon)^2 } + {\epsilon^2 \over 2 ( i \epsilon -\delta^2 {\bf k^2})^2} \right\}
\nonumber \\
&=&-1.
\end{eqnarray} 

\begin{center}
{\bf Acknowledgments}
\end{center}

D.J.Lee is very grateful to I.J.R Aitchison for useful discussions and for editing the manuscript.

\begin{center}
{\bf References}
\end{center}

\begin{tabbing}

\= $[1]$ \  \=  R.D Pisarki, $Phys.$ $Rev.$ {\bf D29} (1984) 2423\\
\\
\> $[2]$ \> T.W Appelquist, M.Bowick, D.Karabali and L.C.R Wijewardhana\\
\> \>  $Phys.$ $Rev.$ {\bf D33} (1986) 3704. \\
 \> \> T.W Appelquist, D.Nash and L.C.R Wijewardhana, $Phys.$ $Lett.$ {\bf 60} (1988) 2575.\\
\\
\> $[3]$ \> M.R Pennington and D.Walsh, $Phys.$ $Lett.$ {\bf B253} (1991) 246.
\\
\\
\> $[4]$ \> E.Dagotto, A.Kocic and J.B Kogut, $Phys.$ $Rev.$ $Lett.$ {\bf 62} (1989) 1083\\
\> \> and $Nucl.$ $Phys.$ {\bf B334} 279.\\

\> $[5]$ \> D.Nash, $Phys.$ $Rev.$ $Lett.$ {\bf 62} (1989) 3024.\\
\\
 \> $[6]$ \> D.Atkinson, P.W Johnson and P.Maris, $Phys.$ $Rev.$ {\bf D42} 
(1990) 602.\\
\\
\> $[7]$ \> P.Maris, $Phys.$ $Rev.$ {\bf D54} (1996) 4049.\\
\\
\> $[8]$ \> K.-I.Kondo, $Phys.$ $Rev.$ {\bf D55} (1997) 7826\\
\\
\> $[9]$ \> I.J.R Aitchison, N.E.Mavromatos, D.McNeil, $Phys.$ $Lett.$ 
{\bf B402} (1997) 154\\
\\
\> $[10]$ \> N.Dorey and N.E.Mavromatos, $Phys.$ $Lett.$ {\bf B226} (1991) 163
\\
\> \>  and $Nucl.$ $Phys.$ {\bf B386} (1992) 614.\\
\\
\> $[11]$ \> I.J.R Aitchison, N.Dorey, M.Klein-Kreisler and N.E Mavromatos,\\
\> \>  $Phys.$ $Lett.$ {\bf B294} (1992) 91.\\
\\
\> $[12]$ \> I.J.R Aitchison and M.Klein-Kreisler, $Phys.$ $Rev.$ {\bf D50} (1993) 1068.\\
 \\
\> $[13]$ \> I.J.R Aitchison, $Z.Phys.$ {\bf C67} (1995) 303.\\
\\
\> $[14]$ \> D.Lee, $Oxford$ $Unversity$ $PrePrint$ OUTP-97-56-P/hep-th/9803115 (1997).\\
\\
\> $[15]$ \> G.Triantaphyllou hep-th/9801245 (1998)\\
\\
\> $[16]$ \>  D.Lee and G.Metikas, $Oxford$ $Unversity$ $PrePrint$ OUTP-97-72-P\\
\> \> /hep-th/9803211 (1997).\\

\end{tabbing}
\newpage
 
\begin{center}
{\bf Table 1}
\end{center}
\begin{table}
\begin{center}
\begin{tabular}{|c|c|}
\hline $N$ & $r$ \\ \hline
1 & 5.59 \\ \hline
2 & 5.57 \\  \hline
3 & 5.51  \\ \hline
\end{tabular} \vskip0.10in
\end{center}
\caption{The value of the quanity $r$ as defined in the text at the indicated values of $N$.} 
\end{table}

\begin{center}
 {\bf Figure Captions}
\end{center}

Figure.1: A diagramic representation of the S.D equation.\\

Figure.2: Graph showing the constant physical mass solutions as functions of $Tk_B/\alpha$ for $N=1,2$ and $3$.\\

Figure.3: Graph showing the variation of $k_BT_c/ \alpha$ with w.r.t $N$.\\

\newpage

\begin{figure}
\begin{center}
\includegraphics{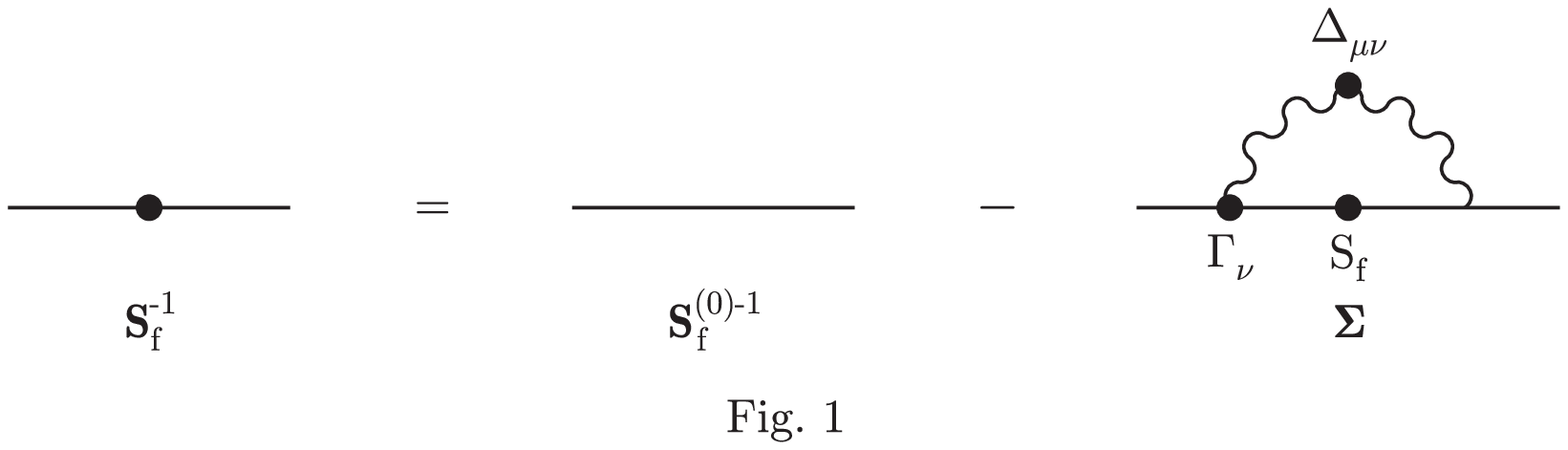}
\end{center}
\end{figure}

\begin{figure}
\begin{center}
\includegraphics{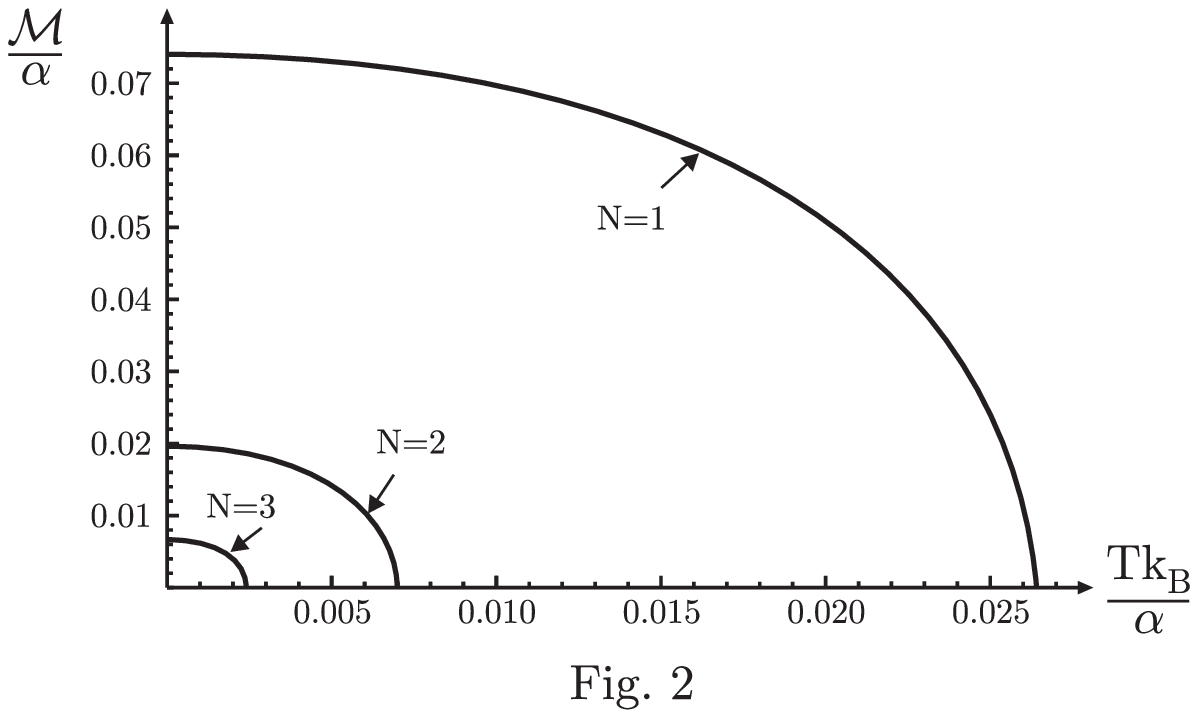}
\end{center}
\end{figure}

\begin{figure}
\begin{center}
\includegraphics{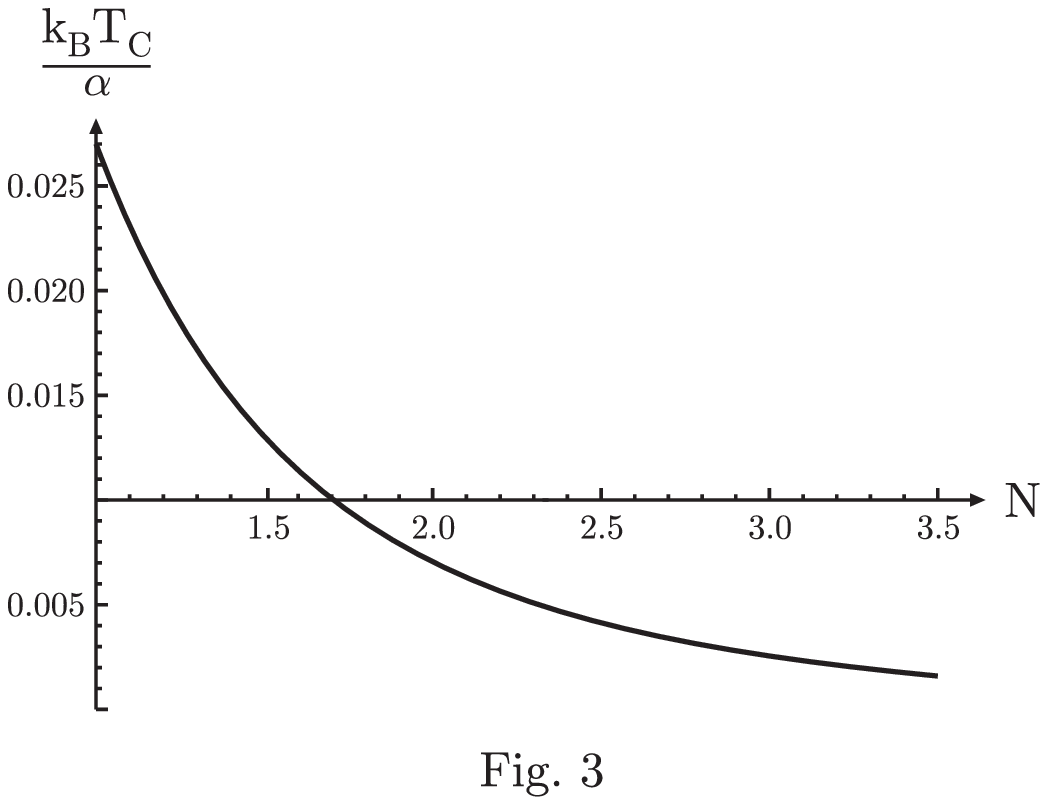}
\end{center}
\end{figure}

\end{document}